\documentclass[lettersize,journal]{IEEEtran}
\usepackage{amsmath,amsfonts}
\usepackage{algorithmic}
\usepackage{algorithm}
\usepackage{array}
\usepackage[caption=false,font=normalsize,labelfont=sf,textfont=sf]{subfig}
\usepackage{textcomp}
\usepackage{stfloats}
\usepackage{url}
\usepackage{verbatim}
\usepackage{graphicx}
\usepackage{cite}
\usepackage{soul}
\usepackage{mathtools}
\usepackage{color,soul}
\usepackage{breqn}
\begin{document}

\newcommand{\e}{\mathcal{E}}
\newcommand{\probP}{\text{I\kern-0.15em P}}

\title{CollabORAN: A Collaborative rApp–xApp–dApp Control Architecture for Fairness-Adaptive Resource Sharing in O-RAN}

\author{Anastasios E. Giannopoulos,~\IEEEmembership{Member,~IEEE,} Sotirios T. Spantideas, and Panagiotis Trakadas

\thanks{This work was partially supported by the UNITY-6G Project, funded by HORIZON-JU-SNS-2024 program, under grant agreement No 101192650. (Corresponding author: Anastasios Giannopoulos.)}%
\thanks{Anastasios Giannopoulos, Sotirios Spantideas and Panagiotis Trakadas are with the Research \& Development Department, Four Dot Infinity (FDI), Chalandri, Athens, P.C. 15231, Greece (e-mail: \{angianno, sospanti, ptrak\} @fourdotinfinity.com).}}%
%\thanks{Digital Object Identifier XX.YYYY/COMMAG.ZZZZ.LLLLLL}}%

\markboth{PREPRINT MANUSCRIPT}%
{Giannopoulos \MakeLowercase{\textit{et al.}}: CollabORAN: A Collaborative rApp–xApp–dApp Control Architecture for Fairness-Adaptive Resource Sharing in O-RAN}

% Remember, if you use this you must call \IEEEpubidadjcol in the second column for its text to clear the IEEEpubid mark.

\maketitle

\begin{abstract}
The evolution of Open Radio Access Networks (O-RAN) enables programmable and intelligent control of radio resources through disaggregated architectures and open interfaces. However, existing solutions typically rely on isolated control loops and fail to jointly address end-to-end optimization objectives across multiple timescales. Thus, it remains a key challenge to functionally split optimization algorithms across timescale-specific O-RAN layers while complying with control loop latency specifications. This article proposes CollabORAN, a collaborative rApp–xApp–dApp hierarchical framework for dynamic and equitable spectrum sharing in O-RAN systems. CollabORAN leverages a nested control structure in which the rApp performs traffic-aware policy generation, the xApp executes interference-aware spectrum allocation via hypergraph-based PRB coloring, and the DU-level dApp enforces temporal fairness through fast scheduling. The proposed end-to-end closed-loop design enables coordinated optimization across minutes, seconds, and millisecond time scales. Simulation results demonstrate that CollabORAN significantly improves service fairness and reduces user starvation while maintaining efficient spectrum reuse in dense and dynamic network environments.
\end{abstract}


\section{Introduction}

\subsection{Modern Open Radio Access Networks}

The rapid evolution of Open Radio Access Networks (O-RAN) is fundamentally reshaping the way radio resources are managed in modern cellular systems \cite{agarwal2025open}. By disaggregating the traditional base station architecture and exposing standardized interfaces, O-RAN enables the deployment of intelligent applications that dynamically optimize network operation. Central to this vision is the RAN Intelligent Controller (RIC), which supports programmable control logic through rApps in the Non-Real-Time RIC (Non-RT RIC) and xApps in the Near-Real-Time RIC (Near-RT RIC). These applications allow operators and developers to introduce policy-driven automation and data-driven optimization into the RAN \cite{polese2023understanding}.

Recent advances in O-RAN architectures are further extending this paradigm with the introduction of distributed applications (dApps) operating closer to the radio access nodes, typically within distributed units (DUs) or Central Units (CUs) \cite{d2022dapps}. While rApps focus on long-term policy definition (seconds-to-hours timescale) and learning-based optimization, and xApps perform near-real-time control (below 1 sec) across multiple Radio Units (RUs), dApps enable ultra-fast decision making (below 10 ms) directly at the RAN execution layer \cite{spantideas2026fedra}. This emerging three-tier intelligence hierarchy (rApp, xApp, and dApp) creates new opportunities for coordinated multi-timescale optimization of radio resources, allowing policies, interference management, and scheduling decisions to be jointly orchestrated across different control layers \cite{d2022dapps}.

At the same time, cellular networks are facing increasing pressure on the radio spectrum due to the rapid growth of connected devices, heterogeneous deployments, and data-intensive services. Dense multi-cell environments introduce complex interference relationships among users and base stations, which significantly complicates spectrum allocation decisions. Efficient spectrum access mechanisms must therefore address several competing objectives simultaneously, including interference mitigation, efficient spectrum reuse, and fair resource distribution among users \cite{anand2026mitigating}.

\subsection{Gaps, Challenges and Contributions}

Existing resource allocation approaches in O-RAN environments typically rely on either static scheduling policies implemented at the base station level or isolated optimization mechanisms implemented within individual xApps \cite{anand2026mitigating, zhang20245g}. Static schedulers are inherently limited in their ability to adapt to dynamic traffic demand, user mobility and interference conditions in dense and heterogeneous deployments. At the same time, most current O-RAN control solutions focus on single-loop optimization performed at the Near-RT RIC, where an xApp attempts to optimize specific network objectives such as interference mitigation, traffic steering, load balancing, or throughput maximization \cite{bonati2021intelligence}.

While xApps enable near-real-time control across multiple cells, relying solely on xApp-level optimization introduces several limitations. First, xApps typically operate on short time horizons and limited network context, which restricts their ability to incorporate long-term policy objectives such as service-level agreements (SLAs), traffic forecasts, or operator-defined fairness targets. Second, xApps generally assume that resource scheduling and enforcement occur locally at the RAN nodes, without explicitly coordinating with execution-layer intelligence. %As a result, optimization decisions made at the Near-RT RIC may not fully translate into efficient resource utilization at the scheduling level. 
Similarly, rApps operating in the Non-RT RIC provide powerful capabilities for long-term policy design and AI-driven network optimization. However, rApps alone cannot react to fast-changing radio conditions or enforce resource allocation decisions at fine time granularity \cite{polese2023understanding}. %This separation between policy definition and real-time execution creates a control gap where high-level optimization objectives cannot be consistently translated into actionable scheduling decisions within the RAN.
In addition, the architectural introduction of dApps in the O-RAN control loops enables direct radio control in sub-slot or even TTI-level timescales, aligning with the latency requirements of scheduling and far-edge radio resource allocation \cite{d2022dapps,spantideas2026fedra}. Hence, the adoption of dApps opens new opportunities for hierarchical control loops in which long-term policies (defined by rApps) and interference-aware resource allocation (performed by xApps) can be complemented by ultra-fast scheduling and resource arbitration mechanisms at the DU level.

Despite these emerging capabilities, existing O-RAN optimization frameworks largely treat rApps, xApps, and execution-layer schedulers at the DU as independent control entities. Coordinating these layers introduces several challenges, including how to distribute optimization tasks across different time scales for end-to-end optimization, how to maintain stability across nested control loops, and how to exchange information between policy, control, and execution layers without introducing excessive signaling overhead. %At the same time, such hierarchical coordination offers significant benefits, including improved responsiveness to dynamic radio conditions, more efficient interference management, and the ability to enforce long-term policy objectives through fine-grained resource scheduling.
Furthermore, relying solely on rApp- or xApp-based control not only concentrates optimization decisions at the Non- or Near-RT RIC, often leading to frequent O1 or E2 signaling and potential interface overload, but also limits decision responsiveness to minute- or second-level timescales \cite{kanthaliya2026load}. In contrast, a coordinated and nested rApp–xApp–dApp architecture can decouple control responsibilities across layers, balance communication overhead, and enable responsive resource management across multiple time scales \cite{d2022dapps}. %Therefore, a key open challenge in intelligent O-RAN design is how to realize collaborative multi-timescale control frameworks that coordinate rApps, xApps, and dApps in a unified optimization workflow. Such frameworks must bridge the gap between policy-driven network management, interference-aware spectrum allocation, and fast decentralized scheduling decisions.

%\subsection{CollabORAN Framework Contributions}

To address this gap, this article proposes CollabORAN, a collaborative spatiotemporal spectrum management framework that enables coordinated rApp–xApp–dApp control loops for equitable spectrum access in O-RAN networks. By decoupling the fair spectrum allocation problem into different timescale-specific tasks, the proposed architecture leverages a hierarchical O-RAN structure to separate an AI/ML-aided and SLA-driven policy definition (through AI/ML traffic forecasting), interference-aware spectrum allocation (through dynamic hypergraph construction and coloring), and resource time-sharing scheduling across different O-RAN layers and time scales. %Specifically, the rApp defines fairness policies and traffic-aware objectives, the xApp constructs dynamically interference conflict hypergraphs (i.e. graphs that take into account both inter-RUs and intra-RU conflicts at once) and performs spectrum allocation through mobility-aware graph coloring, and the dApp deployed at the DU level enforces temporal fairness through modified proportional-fair (PF) scheduling and dynamic resource sharing.
%The proposed CollabORAN enables end-to-end multi-timescale coordination between policy intent, interference-aware spectrum allocation, and temporal fairness enforcement, allowing the network to achieve both efficient spectrum reuse and equitable spectrum access across users. 
The specific contributions of CollabORAN are summarized as follows:

\begin{itemize}
    \item The framework provides generalized collaborative multi-timescale rApp–xApp–dApp control that enables end-to-end policy-driven dynamic spectrum management in O-RAN networks. Policies at the Non-RT RIC are dynamically co-configured by the operator-defined SLAs and AI/ML-assisted traffic forecasts from the rApp (called FrerApp). 
    \item A graph-based interference-aware and mobility-aware physical resource block (PRB) allocation mechanism (called FrexApp) is implemented at the xApp layer to dynamically construct interference conflict graphs and allocate spectrum resources across users covered by heterogeneous RUs.
    \item A graph-constrained modified proportional-fair (PF) temporal scheduling mechanism (called FredApp) is applied at the dApp layer, enabling dynamic PRB time sharing that prevents long-term user starvation and improves fairness.
    \item A comprehensive and generalizable workflow analysis of the proposed hierarchical and collaborative closed loop is presented. In addition, performance evaluation in heterogeneous O-RAN environments is conducted to demonstrate the CollabORAN performance in terms of traffic forecasting, interference mitigation and service share balance, under dynamic traffic conditions. The results are compared against standalone deployments of intelligent applications to showcase the tangible potency of CollabORAN.
\end{itemize}

\section{CollabORAN Architectural Overview}

\subsection{Design Principles}

The design of CollabORAN is guided by three key architectural principles that enable scalable and coordinated spectrum management across the O-RAN ecosystem.

\textbf{Principle 1 - Separation of concerns}: CollabORAN distributes spectrum management responsibilities across the Non-RT RIC, Near-RT RIC, and RAN execution layers (DU/RU). Each layer performs optimization tasks that match its available network visibility and operational timescale. This separation avoids overloading a single control entity and enables each component to operate using the most appropriate network information and control granularity.

\textbf{Principle 2 - Policy-Control-Execution hierarchy}: The framework follows a hierarchical control pipeline in which high-level policies are defined and updated at the Non-RT RIC, interference-aware spectrum allocation decisions are performed at the Near-RT RIC, and fast resource scheduling is executed directly within the RAN nodes. This hierarchical structure ensures that long-term operator objectives, such as priorities, fairness policies and SLAs, can be translated into actionable resource allocation decisions that are enforced at the scheduling level.

\textbf{Principle 3 - Timescale separation}: Spectrum management decisions are executed across multiple time scales. Long-term policy configuration and traffic forecasting operate at the Non-RT RIC on a minutes-to-hours horizon. Near-real-time spectrum allocation decisions at the Near-RT RIC are performed on the order of seconds. Finally, execution-layer scheduling decisions at the DU occur at millisecond-level timescales. This temporal separation allows CollabORAN to combine strategic optimization with fast reaction to dynamic network conditions.

\subsection{CollabORAN Functional Architecture}

\begin{figure}[t]
\centering
\includegraphics[trim={0.6cm 0.8cm 0.4cm 0.6cm},clip,width=1\columnwidth]{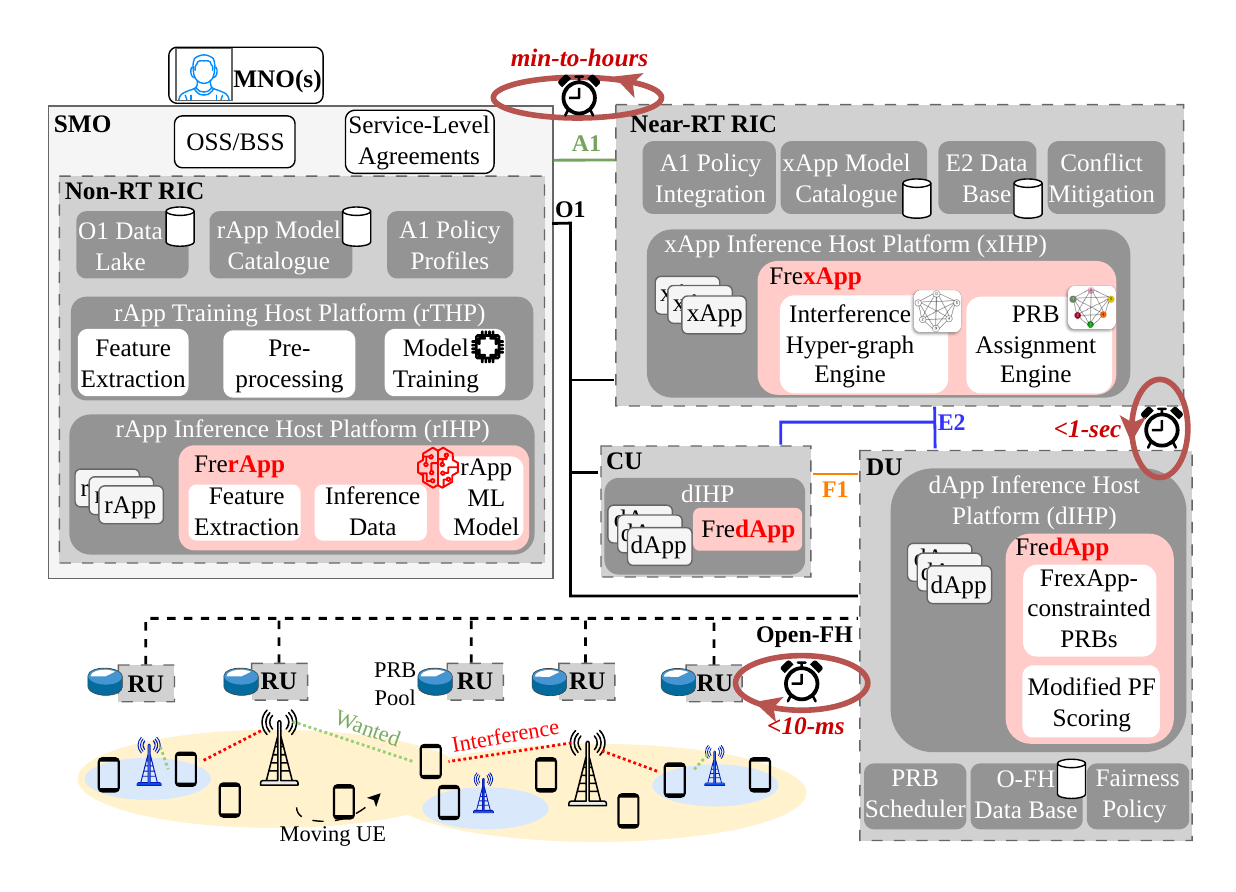}
\caption{CollabORAN layered architecture enabling hierarchical multi-timescale spectrum sharing through collaborative rApp–xApp–dApp control interactions.}
\label{fig:fig1}
\end{figure}

The overall architecture of CollabORAN is shown in Fig.~\ref{fig:fig1}. The framework integrates components across the \textit{Service Management and Orchestration (SMO)} layer, the \textit{Non-RT RIC}, the \textit{Near-RT RIC}, and the \textit{RAN execution layer} (CU/DU and RUs). Each layer contributes distinct functionalities to support policy-driven and interference-aware spectrum management across multiple timescales \cite{garcia2021ran}.

At the SMO layer, the \textit{Mobile Network Operator (MNO)} defines high-level operational objectives and service requirements. These requirements (e.g. UE priorities, throughput satisfaction tolerance) are typically expressed through SLAs and operational policies derived from \textit{Operational/Business Support Systems systems (OSS/BSS)}. The SMO acts as the entry point for operator-defined policies and long-term network objectives, which are translated into machine-readable policy profiles and propagated toward the RIC layers through standardized O-RAN management interfaces.

The Non-RT RIC provides long-term intelligence and data-driven network optimization capabilities. This layer aggregates network measurements and historical operational data through the O1 interface, which feeds the \textit{O1 Data Lake} used for data analytics and model training. The Non-RT RIC hosts the lifecycle management of AI/ML models through dedicated components such as the \textit{Model Catalogue} and the \textit{rApp Training Host Platform (rTHP)}, where functions including feature extraction, data preprocessing, and model training are performed. Once trained, models are deployed to the \textit{rApp Inference Host Platform (rIHP)}, where they process incoming network data to generate high-level control guidance and policy updates. These outputs combine a set of SLA-defined and traffic prediction-assisted policies to form a dynamic \textit{Policy Profile}, which is communicated to the Near-RT RIC through the A1 interface, enabling the translation of long-term policies into actionable control directives of xApps.

The Near-RT RIC performs near-real-time network control functions using information received from both the Non-RT RIC and the E2 RAN execution layer. Within this layer, the \textit{A1 Policy Integration} module receives policy directives originating from the SMO and Non-RT RIC. The Near-RT RIC also maintains an \textit{xApp Model Catalogue} and an \textit{E2 Data Base}, which store operational measurements and network state information collected through the E2 interface from the CU/DU nodes. These datasets are processed by the \textit{xApp Inference Host Platform (xIHP)}, which executes near-real-time control applications responsible for translating policies and applying optimization algorithms into concrete radio resource management decisions. The \textit{Conflict Mitigation} module is in charge of identifying and resolving conflicts between running xApps. In the context of spectrum management, xIHP includes modules for dynamic interference hypergraph construction and PRB allocation, enabling coordinated resource allocation and scheduling across multiple RUs.

The RAN execution layer, composed of the CU/DU platforms, is responsible for enforcing the control decisions produced by the xApp. Communication between the Near-RT RIC and the RAN nodes is performed through the E2 interface, while coordination between CU and DU entities is handled via the F1 interface. Within the DU platform, the \textit{dApp Inference Host Platform (dIHP)} executes DU control functions that interact directly with the MAC PRB scheduler and local RAN databases. These components enable fast resource scheduling decisions that respond to short-term channel variations and traffic dynamics. Although dApps can alternatively be hosted in the CU, here we focus on dApps at the DU to take advantage of the MAC scheduler co-location and avoid extra CU-DU communication overhead during the PRB scheduling.

At the bottom of the architecture, the O-RAN infrastructure consists of multiple RUs connected with the DU through the Open Fronthaul (Open-FH) interface. The RUs serve mobile user equipments (UEs) and share a common pool of available PRBs. Due to user mobility, dynamic traffic, and heterogeneous deployment scenarios (e.g. macro-, micro-, pico-cells or aerial cells), transmissions from different RUs may interfere when the same spectrum resources are reused simultaneously. The PRB pool must therefore be dynamically managed to balance spectrum reuse and interference mitigation while maintaining service quality and fairness across users.

Through this layered architecture, CollabORAN establishes a coordinated hierarchical control pipeline where operator policies, network analytics and prognostics, near-real-time decision making, and execution-layer scheduling interact through standardized O-RAN interfaces. %Using this architecture, a scalable and responsive spectrum management can be established by distributing control responsibilities across rApps, xApps and dApps.

\subsection{CollabORAN Nested Control}

\begin{figure}[t]
\centering
\includegraphics[trim={0.4cm 1.2cm 0.8cm 0.8cm},clip,width=1\columnwidth]{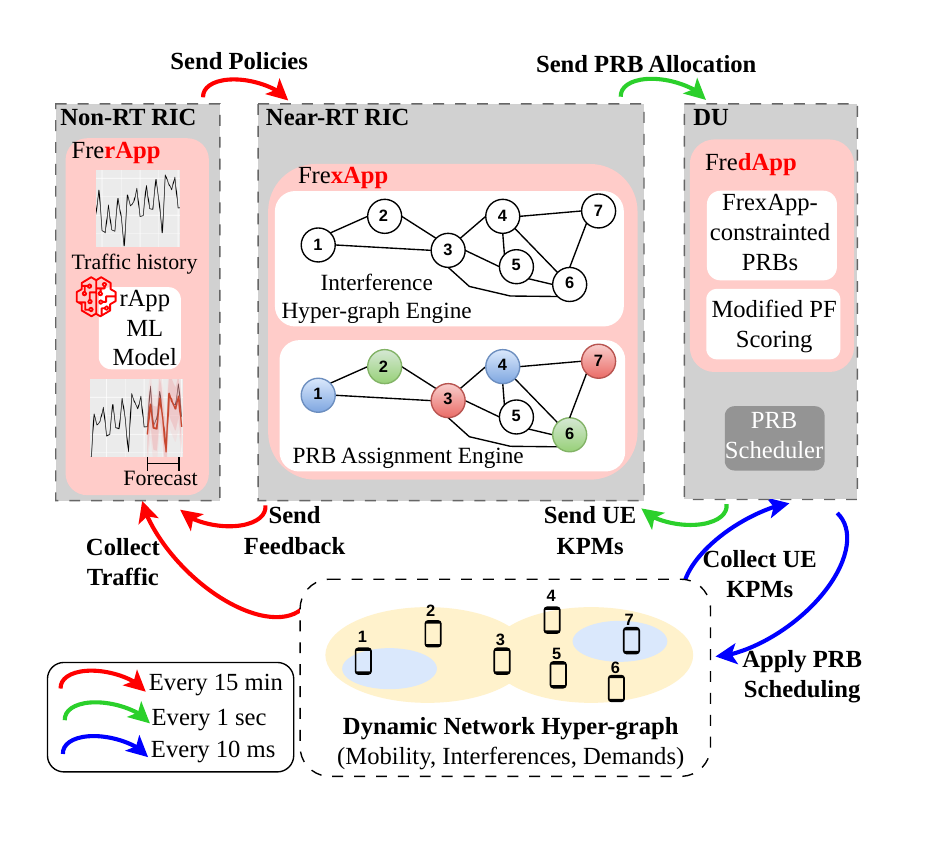}
\caption{Monitoring and control interactions among heterogeneous O-RAN areas, FrerApp, FrexApp and FredApp.}
\label{fig:fig2}
\end{figure}

CollabORAN implements a nested multi-timescale control workflow, depicted in Fig.~\ref{fig:fig2} in which the fair frequency resource allocation process is decomposed across distinct Non-RT RIC, Near-RT RIC and DU applications called FrerApp, FrexApp, and FredApp, respectively. The collaboration among these functions implements the end-to-end spectrum management logic of CollabORAN.

\subsubsection{FrerApp for Policy Definition and Traffic-Aware Control}

The FrerApp is responsible for defining high-level spectrum management policies based on operator objectives and network conditions. It runs at a long timescale ranging from minutes to hours and relies on historical traffic measurements collected through the O1 interface. Using these measurements, the FrerApp performs AI/ML-based traffic forecasting to estimate the expected traffic demand of each RU over a future horizon. Based on these rApp forecasts and operator-defined service-level agreements, the FrerApp generates a policy profile that is acknowledged to the Near-RT RIC. % This policy profile is transmitted to the Near-RT RIC through the A1 interface to steer the behavior of the spectrum management functions. These directives include slice or user priority classes derived from SLAs (e.g. High vs Low), the selection of the fairness policy used for spectrum time-sharing (e.g. PF or Round Robin), interference tolerance levels associated with each user (e.g. 5\% acceptable degradation relative to the requested service), and the choice of PRB assignment algorithm (e.g. rule-based like 'best SINR first', graph coloring, DRL spectrum allocation model). In addition, the FrerApp dynamically determines the spectrum segmentation configuration (5G NR numerology), which defines the subcarrier spacing, the number of available PRBs and their bandwidth according to the predicted traffic intensity and the system channel bandwidth.

\subsubsection{FrexApp for Interference Hypergraph Construction and Spectrum Allocation}

The FrexApp performs network-wide (handles tens of RUs) spectrum coordination at a sub-second timescale. Its primary role is to integrate the individual policies of the policy profile (from the FrerApp) into the xApp logical modules. To achieve this, the FrexApp dynamically constructs an interference hypergraph representing the relationships among all active users in the heterogeneous O-RAN. In the hypergraph, each node represents a user, while edges capture interference conflicts among users that cannot simultaneously reuse the same PRB. Unlike conventional interference graphs that only capture pairwise conflicts within a cell, the hypergraph formulation allows the FrexApp to jointly model both intra-RU and inter-RU interference relationships within a unified representation. The hypergraph is updated every time the FrexApp is triggered (second-level update), allowing the spectrum allocation to adapt to user mobility, traffic fluctuations, and evolving interference conditions.

%The hypergraph is constructed using instantaneous radio measurements collected through the E2 interface, including channel conditions, traffic demand levels, and user-specific interference tolerance margins defined by the policy profile. Based on this dynamic hypergraph representation, the FrexApp performs PRB allocation using a graph-coloring strategy. In this formulation, each color represents a PRB, and users connected through interference relationships cannot be assigned the same color. The resulting coloring determines which users across different radio units can safely reuse the same spectrum resources while maintaining their service satisfaction conditions.

 %The output of the FrexApp is a PRB allocation map that specifies the feasible spectrum assignments for each user while respecting interference constraints.

\subsubsection{FredApp for Fairness-aware Resource Time-Sharing}

While the FrexApp determines which PRBs can be safely reused across the network, some users may remain unassigned when strong interference conditions prevent simultaneous spectrum reuse. To ensure that these users are not permanently excluded from spectrum access, the FredApp enforces temporal resource sharing. It runs at a fast scheduling timescale (below 10 ms) and receives the PRB allocation constraints produced by the FrexApp. To save time, instead of recomputing interference relationships, the FredApp focuses on scheduling users over time while respecting the PRB compatibility constraints defined by the FrexApp-generated hypergraph coloring process. In particular, it enables time-sharing of PRBs among users that cannot simultaneously access the same spectrum resources due to interference conflicts.

%By operating directly at the DU platform, the FredApp can react rapidly to short-term channel variations and traffic fluctuations. Its scheduling decisions ensure that users excluded from a given PRB allocation cycle can still obtain periodic access to spectrum resources over time. This execution-layer scheduling mechanism therefore complements the spatial interference mitigation performed by the FrexApp by introducing a temporal dimension to spectrum sharing in the millisecond scale.

\section{Closed-Loop Control for Dynamic Fair Spectrum Sharing}

\begin{figure}[t]
\centering
\includegraphics[trim={0.3cm 0.6cm 0.3cm 0.2cm},clip,width=1\columnwidth]{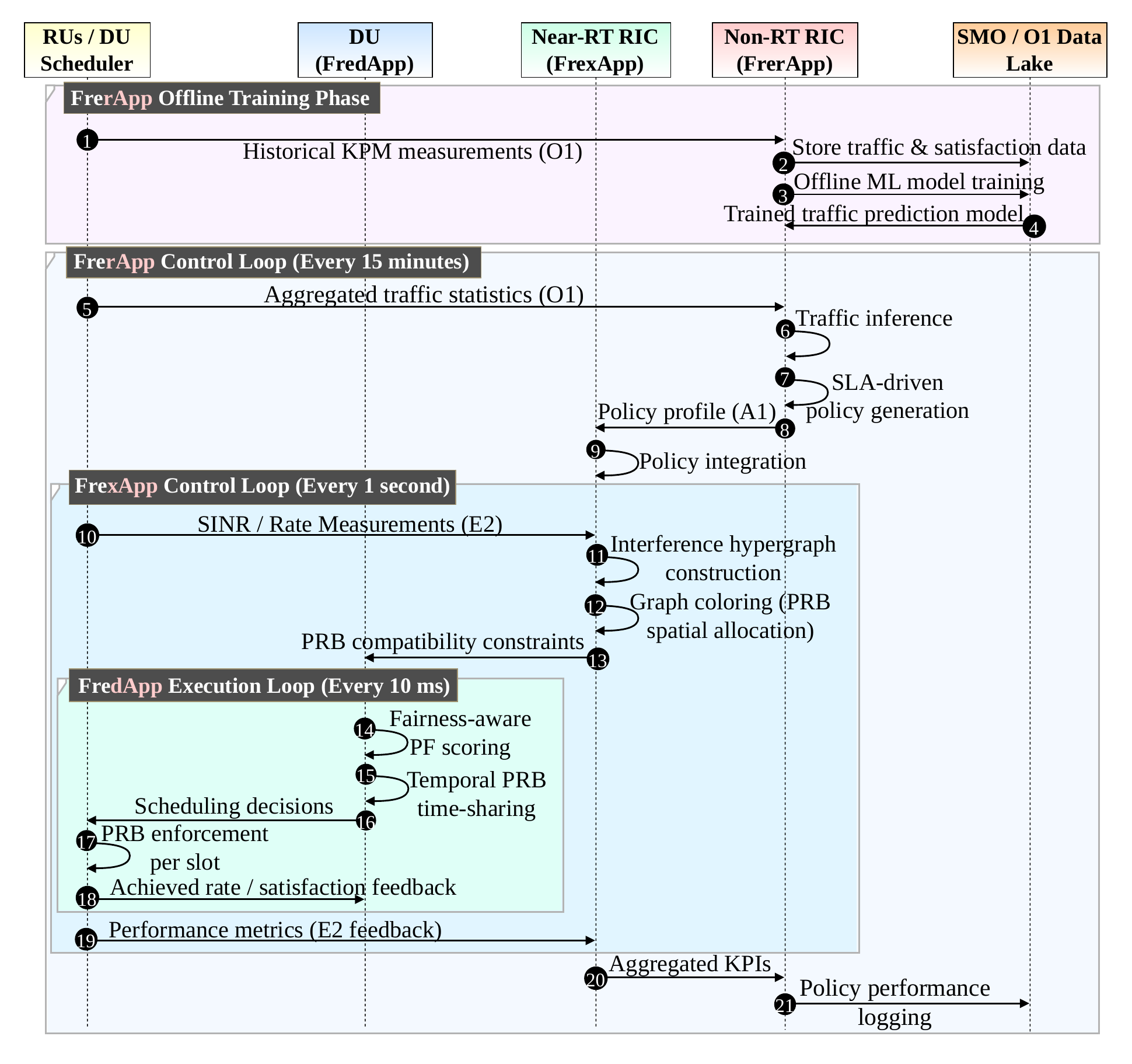}
\caption{CollabORAN Sequence diagram for nested closed-loop control.}
\label{fig:fig3}
\end{figure}

This section describes the operational and analytic workflow of the proposed CollabORAN framework. Fig.~\ref{fig:fig3} illustrates the overall sequence diagram, highlighting the interactions among the SMO, Non-RT RIC, Near-RT RIC, and DU components. The workflow consists of three main phases: an offline training phase, an online rApp control loop, and a nested xApp–dApp control loop, which together enable continuous and adaptive spectrum management.

\textbf{Offline FrerApp Training Phase:} The operation of CollabORAN begins with an offline training phase executed by the FrerApp at the Non-RT RIC. The process starts with the collection of historical key performance measurements from the RAN infrastructure (\textit{Step~1}). These measurements, which include aggregated traffic load per RU, user density, mobility statistics, and long-term service satisfaction indicators, are forwarded via the O1 interface and stored in the SMO data repository (\textit{Step~2}), forming a historical dataset that captures the traffic evolution of the network. Using this dataset, the FrerApp performs offline training of a traffic prediction model (\textit{Step~3}). The objective of this process is to learn temporal traffic patterns and estimate the maximum expected UE density across RUs during upcoming control intervals. Since this training is executed offline, it does not affect the responsiveness of the real-time RAN operation. Once training is completed, the resulting traffic prediction model is deployed within the FrerApp inference environment (\textit{Step~4}), where it can later be used to generate traffic forecasts and policy configurations. Periodic retraining may also be performed as new traffic traces become available in the data repository.

\textbf{Online FrerApp Control Loop:} After the traffic prediction model has been trained, the FrerApp executes an online control loop operating at a minutes-level timescale (e.g., every 15 minutes). At the beginning of each cycle, the FrerApp receives updated aggregated traffic statistics through the O1 interface (\textit{Step~5}), which reflect the current load conditions across the RU nodes. Using these measurements, the FrerApp performs traffic inference based on the previously trained model (\textit{Step~6}) in order to estimate the expected peak UE density across RUs during the next control horizon. These predictions enable the system to anticipate potential congestion situations and adjust the spectrum management strategy accordingly. Based on the predicted traffic conditions and the operator-defined SLAs, the FrerApp then constructs a policy profile that defines the configuration parameters for the Near-RT spectrum management functions (\textit{Step~7}). The generated policy profile is transmitted to the Near-RT RIC through the A1 interface (\textit{Step~8}). Importantly, the FrerApp does not directly allocate spectrum resources. Instead, it configures the operational parameters and optimization strategies used by the Near-RT control modules. This policy profile includes a set of policies $\boldsymbol{{\pi}}=\{\pi_1,\pi_2,\pi_3,\pi_4,\pi_5\}$, where $\pi_1$ represents the user priority classes, $\pi_2$ denotes the degradation tolerance levels, $\pi_3$ defines the selected PRB allocation strategy, $\pi_4$ sets the fairness scheme, and $\pi_5$ justifies the spectrum segmentation configuration (5G NR numerology) and is set dynamically based on the FrerApp traffic predictions. Upon reception of this policy configuration, the FrexApp integrates the received parameters into its internal spectrum allocation engines (\textit{Step~9}).

\textbf{Nested FrexApp–FredApp Control Loop:} The Near-RT RIC and DU components form a nested control loop responsible for spectrum allocation and scheduling execution. In this hierarchical structure, the FrexApp operates at a near-real-time timescale (approximately every second), while the FredApp executes a faster scheduling loop at the DU level every 10~ms. At the beginning of the FrexApp control cycle, the Near-RT RIC collects instantaneous radio measurements from the RAN nodes through the E2 interface (\textit{Step~10}), including SINR values, achievable data rates, and other channel quality indicators for all active users over a reference PRB. Based on these measurements and the policy parameters provided by the FrerApp, the FrexApp dynamically constructs an interference hypergraph representing the relationships among active users (\textit{Step~11}). In this representation, each node corresponds to a UE, while edges capture interference relationships indicating that two users cannot simultaneously occupy the same PRB without violating their performance requirements. Using this hypergraph representation, the FrexApp performs spectrum allocation through a graph-coloring procedure (\textit{Step~12}), where each color corresponds to a PRB and the coloring constraints ensure that interfering users are not assigned the same PRB. The resulting coloring solution defines PRB compatibility constraints across users and cells (\textit{Step~13}), which are then forwarded to the DU-level scheduling entity.

At the DU, the FredApp executes the fast execution-layer scheduling loop. First, the FredApp evaluates users using a conflict-aware proportional-fair scoring mechanism (\textit{Step~14}), which accounts for user priorities, inter-UE graph conflicts and historical throughput performance. Based on this metric, the scheduler performs temporal PRB time-sharing among compatible users (\textit{Step~15}), allowing users that remain temporarily unassigned after the coloring phase to opportunistically access spectrum resources. The resulting scheduling decisions are then produced and forwarded to the DU scheduler (\textit{Step~16}), which enforces the PRB assignments at the slot level across the connected RU nodes (\textit{Step~17}). Following scheduling execution, the achieved user data rates and satisfaction indicators are monitored (\textit{Step~18}), and the corresponding performance metrics are reported back to the Near-RT RIC through the E2 interface (\textit{Step~19}). The Near-RT RIC aggregates these metrics into higher-level KPIs (\textit{Step~20}), which are subsequently forwarded to the Non-RT RIC where policy performance information is logged in the SMO data repository (\textit{Step~21}). This feedback mechanism closes the hierarchical control loop, allowing CollabORAN to continuously adapt its traffic prediction models, spectrum allocation strategies, and scheduling policies according to evolving network conditions.

\section{Performance Evaluation}

This section demonstrates the performance of the CollabORAN framework in a dynamic multi-cell O-RAN environment. We focus on illustrating the behavior of the multi-timescale control loops and their impact on user satisfaction and fairness.

\subsection{Radio Environment Configuration}

We consider a heterogeneous two-cell O-RAN deployment with a Macro and a Micro-RU, serving mobile users operating over a shared 5G NR spectrum of $10$-MHz bandwidth at $3.5$ GHz frequency. The macro RU operates at higher transmit power ($20$ dBm) to reflect its wider coverage area, while micro RUs provide localized coverage with a $10$-dBm power level. The system operates under the CollabORAN hierarchical control structure, where the FrerApp executes at a $15$-minute timescale, the FrexApp performs spectrum allocation through interference hypergraph coloring at $1$-second intervals, and the FredApp executes a modified PF-driven scheduling decisions \cite{shah2025proportional} at the $10$-ms level. Each rApp control interval consists of $900$ xApp updates, and each xApp interval includes $100$ DU scheduling slots. User positions evolve over time according to a random-walk mobility model, leading to dynamic interference conditions and a time-varying interference hypergraph. Pathloss coefficients, channel models, and SINR calculations for UE reception are derived as in \cite{giannopoulos2025moving}.

To emulate realistic traffic dynamics, we use a public 5G traffic dataset \cite{sevgican2020intelligent}, which provides time-series measurements sampled at 15-minute intervals. The traffic load per RU is mapped to the number of active users and their aggregate demand, allowing the FrerApp to capture cell-level traffic variations. The dataset is used to train and evaluate the FrerApp's traffic prediction model (LSTM regressor), enabling the system to dynamically select the spectrum segmentation policies (by adjusting the 5G NR numerology $\mu=\{0,1,2,3,4\}$) according to anticipated load conditions. Users are assumed to have equal priority and service requirements (requested throughput of $2$ Mbps), focusing the evaluation on the ability of the proposed scheme to manage interference conflicts (by the FrexApp) and ensure equitable spectrum access (by the FredApp) under dynamic network conditions.

\begin{figure*}[t]
\centering
\includegraphics[trim={0cm 1.2cm 0cm 0.7cm},clip,width=1.7\columnwidth]{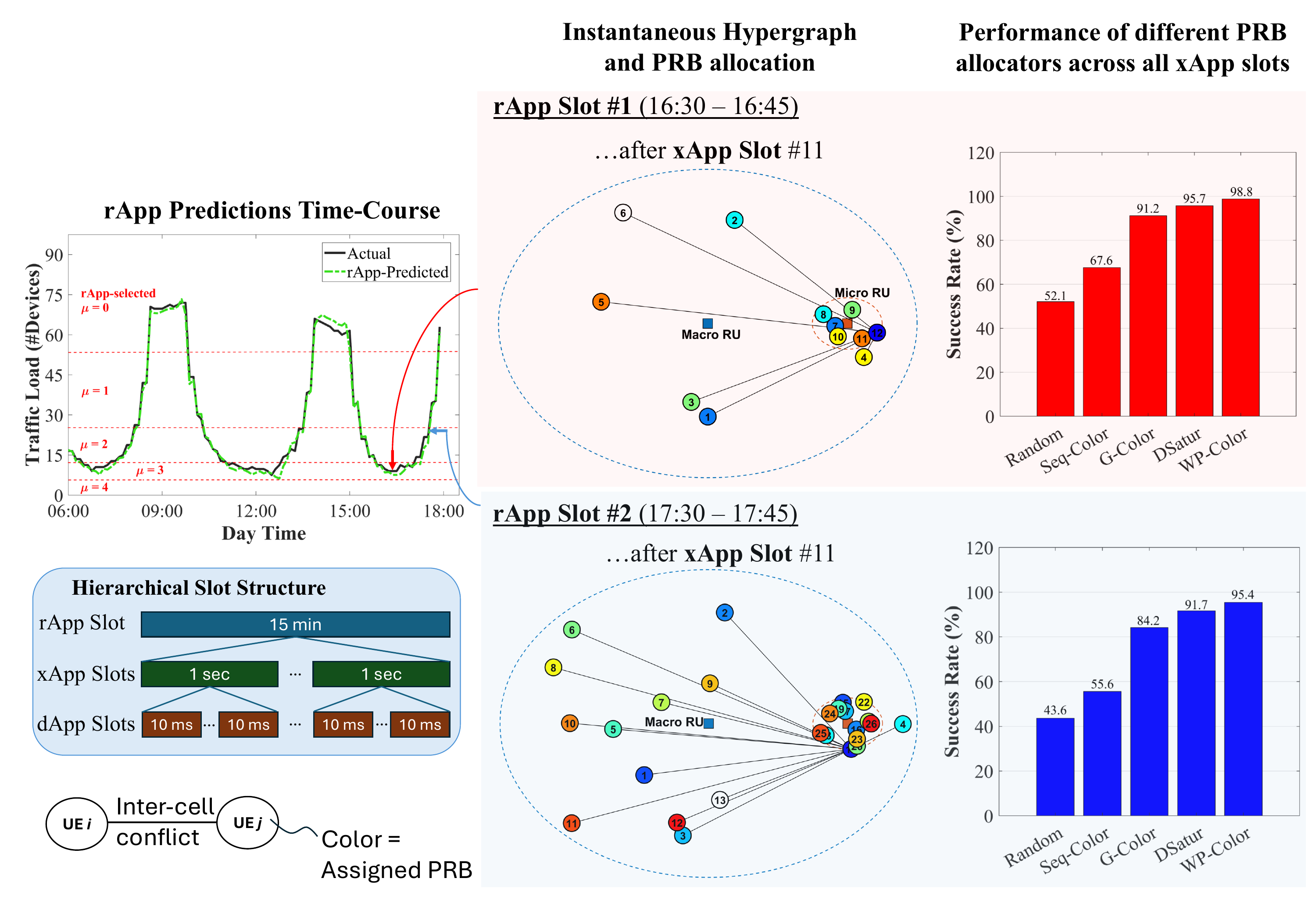}
\caption{Time-evolving behavior of CollabORAN across rApp/xApp control loops. \textbf{Left:} FrerApp traffic prediction versus actual load (dashed red lines define the numerology selection areas). \textbf{Middle:} Instantaneous interference hypergraphs and PRB allocations (different UE colors) produced by the FrexApp during the first xApp slot of two representative rApp intervals (low-load with 12 UEs and higher load with 26 UEs). \textbf{Right:} Average success rate (percentage of satisfied UEs) across all xApp slots for different PRB allocation strategies.}
\label{fig:fig4}
\end{figure*}

\subsection{Time-Evolving rApp/xApp Behavior}

Fig.~\ref{fig:fig4} illustrates the dynamic behavior of the proposed CollabORAN framework across the rApp and xApp control loops. The left panel presents the time evolution of the traffic load along with the corresponding FrerApp predictions over a representative period of a day. The predicted traffic closely follows the actual load variations, enabling the rApp to dynamically select the appropriate spectrum segmentation (numerology) every 15-minute interval. Two representative rApp slots are highlighted, corresponding to different traffic regimes and numerology configurations (rApp slot \#1 corresponds to 16:30-16:45 interval, rApp slot \#2 to 17:15-17:50).

The middle panels depict the instantaneous interference hypergraph and PRB allocation obtained by the FrexApp during the eleventh (selected indicatively) xApp slot of two different rApp intervals. In the first case (rApp Slot~\#1), the network is lightly loaded with 12 users, resulting in a relatively sparse hypergraph and more flexible PRB reuse. In contrast, in the second case (rApp Slot~\#2), the number of users increases to 26, leading to a denser hypergraph with more interference constraints. For clarity, only inter-cell conflicts are visualized, although intra-cell conflicts are also considered in the allocation process. In both cases, the graph-coloring process assigns PRBs (different colors) to a subset of users while leaving some users temporarily unassigned due to conflict constraints. These unsatisfied users are then handled by the FredApp through temporal PRB time-sharing within the same xApp interval, ensuring that they are not persistently excluded from spectrum access.

The right panels present the average success rate (defined as the percentage of satisfied users), across all xApp slots within each rApp interval, for different PRB allocation strategies implemented at the FrexApp. The results show that graph-based approaches significantly outperform baseline methods such as \textit{Random} (each PRB is assigned randomly to a UE) and sequential allocation (\textit{Seq-Color}, PRBs are assigned sequentially to UEs, i.e., PRB 1 to UE 1, and so on). Heuristic graph coloring schemes such as Greedy Coloring (\textit{G-Color}) \cite{namooltree2015solving}, Degree of Saturation (\textit{DSatur}) \cite{san2012new} and Welsh-Powell-based coloring (\textit{WP-Color}) \cite{namooltree2015solving} exhibit success rates above 84\%. In particular, the Welsh-Powell-based coloring achieves the highest success rate in both traffic conditions, reaching up to 98.8\% in low-load scenarios and maintaining strong performance (95.4\% success) under higher load.

\subsection{Comparative Evaluation}

\begin{figure}[t]
\centering
\includegraphics[trim={0cm 0cm 0cm 0cm},clip,width=1\columnwidth]{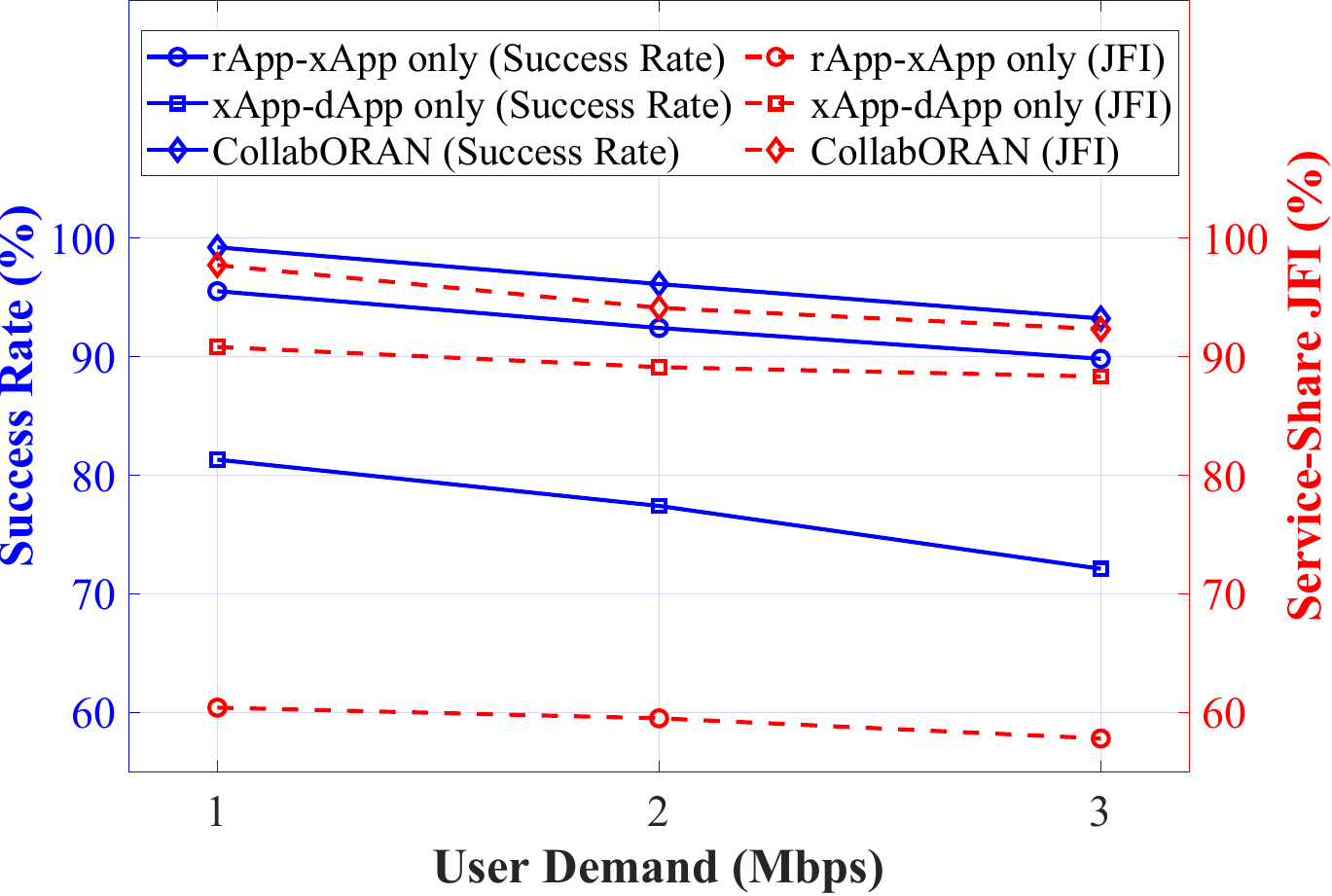}
\caption{Success rate (left y-axis, blue lines) and service-share JFI (right y-axis, red lines) versus user demand for different collaborative schemes.}
\label{fig:fig5}
\end{figure}

Fig.~\ref{fig:fig5} compares the performance of three collaborative control strategies, namely the (i) rApp–xApp only, (ii) xApp–dApp only, and (iii) the CollabORAN framework (combined rApp–xApp–dApp control), under increasing user demand. The evaluation is conducted over a full day of traffic variations, reporting both the success rate (left y-axis) and the service-share fairness in terms of Jain’s Fairness Index (JFI) (right y-axis). For each user, the service-share denotes the fraction of scheduling intervals in which user is fully satisfied.

Evidently, as user demand increases from 1 to 3 Mbps, all schemes experience a slight performance degradation due to intensified competition for limited PRBs. However, clear differences emerge across the three approaches. The \textit{rApp–xApp-only} scheme achieves relatively high success rates, exceeding 90\% even at high demand, but exhibits significantly lower fairness (around 60\%), indicating that a subset of users dominates resource allocation while others remain under-served. This highlights the limitation of relying solely on spatial allocation without fine-grained PRB time-sharing.

In contrast, the \textit{xApp–dApp-only} scheme substantially improves fairness, achieving JFI values above 88\% across all demand levels due to the introduction of dApp-level time-sharing mechanisms. However, this comes at the cost of reduced success rate, which drops below 75\% at higher demand levels, since the absence of long-term policy guidance leads to suboptimal spectrum configuration decisions.

The proposed CollabORAN framework consistently achieves the best trade-off between user satisfaction and fairness. It maintains near-optimal success rates (above 91\%) while simultaneously achieving near-optimal fairness (above 92\%) across all demand levels. This performance stems from the coordinated operation of the multi-timescale control loops, where the FrerApp provides long-term policy guidance, the FrexApp ensures interference-aware spatial allocation, and the FredApp enforces fairness through temporal PRB sharing. Overall, Fig.~\ref{fig:fig5} demonstrates that the hierarchical interplay among rApp, xApp and dApp effectively balances service efficiency and equitable resource distribution, highlighting the necessity of jointly exploiting multi-timescale and multi-layer functionalities for fairness-adaptive resource sharing in O-RAN.

\subsection{Considerations on Generalized Algorithmic Split in O-RAN}

Although CollabORAN is demonstrated for the spectrum allocation optimization, it is worth noting that it can be easily extended to additional integral problems (e.g., minimization of the total network energy efficiency, increased user QoS, etc.), since different intelligent applications deployed at different layers contribute to the same optimization objective. In this context, we showcase how the intelligent O-RAN control loops can be encapsulated to work cooperatively instead of relying only in their standalone operation. Thus, the hierarchically lower dApp uses the intelligent outputs that originate from the xApp which, in turn, incorporates the policy received by the rApp towards the same optimization objective. This cooperative deployment enables the decomposition of a general AI/ML algorithm into multiple intelligent components (functional \textit{algorithmic split}), each fitting into the relevant control loops \cite{matoussi20205g}. Different parts of an end-to-end single-purpose AI/ML algorithm can be therefore executed at different time scales according to the functionality of the hosting network components (e.g., intelligent scheduling in the DU, radio control in the Near-RT RIC, network management decisions in the Non-RT RIC), conforming with the O-RAN requirements and reducing the latency of the control data transfer compared to more centralized architectures of the intelligent components. 

\section{Conclusion}

This article introduced CollabORAN, a multi-timescale collaborative control framework for dynamic and fair resource sharing in O-RAN. By functionally decomposing spectrum optimization across the Non-RT RIC (FrerApp to provide long-term traffic-aware policy configuration), Near-RT RIC (FrexApp to perform interference-aware spatial PRB allocation through hypergraph-based modeling), and DU (FredApp to ensure short-term fairness via fast time-sharing mechanisms) layers, the proposed architecture enables coordinated operation of rApp, xApp, and dApp closed-loop control, each aligned with its respective latency and scope. In summary, CollabORAN can provide a practical and scalable pathway toward intelligent, fair, and adaptive control in O-RAN, addressing a key challenge in the functional split of AI-driven RAN optimization.

\bibliographystyle{IEEEtran}

\bibliography{bibliography}

% \newpage

%\vspace{5pt}

\begin{IEEEbiography}[{\includegraphics[width=1in,height=1.25in]{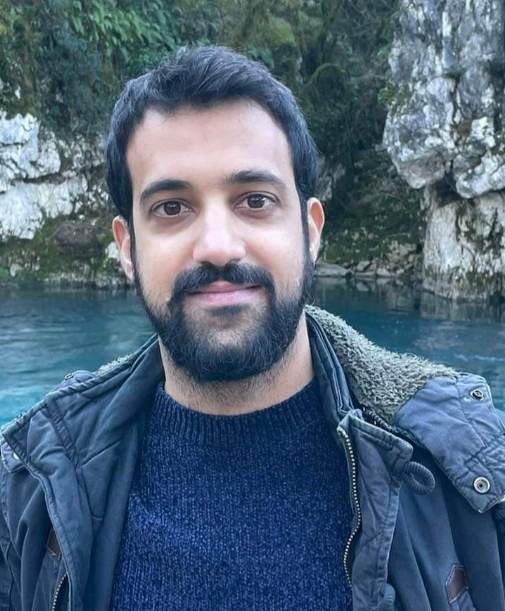}}]{ANASTASIOS E. GIANNOPOULOS}~(Member, IEEE) (M.Eng, Ph.D) received the diploma of Electrical and Computer Engineering from the National Technical University of Athens (NTUA), where he also completed his Master Engineering (M.Eng) degree, in 2018. He also obtained his Ph.D. at the Wireless and Long Distance Communications Laboratory of NTUA. His research interests include advanced Optimization Techniques for Wireless Systems, ML-assisted Resource Allocation, Maritime Communications and Multi-dimensional Data Analysis.

He is currently working as a Research Associate at Department of Ports Management and Shipping, in the National and Kapodistrian University of Athens, as well as Post-Doc at NTUA. He has authored more than 38 scientific publications in the fields of Wireless Network Optimization, Maritime Communications, Machine Learning and Brain Multi-dimensional Analysis. Since 2022, he is a Member of IEEE and reviewer in several IEEE journals (IEEE Internet of Things Journal, IEEE Vehicular Technology Magazine, IEEE Network, IEEE Access).
\end{IEEEbiography}

\begin{IEEEbiography}[{\includegraphics[width=1in,height=1.25in]{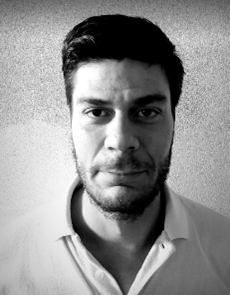}}]{SOTIRIOS T. SPANTIDEAS}~(D.Eng, M.Sc, Ph.D) obtained the Diploma of Electrical $\&$ Computer Engineering from the Polytechnic School of the University of Patras in 2010. He then attended the Master Program "Electrophysics" at the Royal Institute of Technology in Stockholm (KTH), from which he obtained the title MSc in 2013. In 2018 he obtained his PhD from the National Technical University of Athens (NTUA) with doctoral dissertation entitled ``Development of Methods for obtaining DC and low frequency AC magnetic cleanliness in space missions". His research interests include Electromagnetic Compatibility, Machine Learning for Wireless Networks, Magnetic Cleanliness for space missions and optimization algorithms for Computational Electromagnetics.

From 2014, he is working as a Research Associate with NTUA and the National and Kapodistrian University of Athens (Department of Ports Management and Shipping - NKUA), participating in multiple Horizon projects. He has published over 35 papers in scientific journals and conferences in the fields of Electromagnetic Compatibility, Optimization Methods for Wireless Networks and Machine Learning for Resource Allocation problems.
\end{IEEEbiography}

\begin{IEEEbiography}[{\includegraphics[width=1in,height=1.25in]{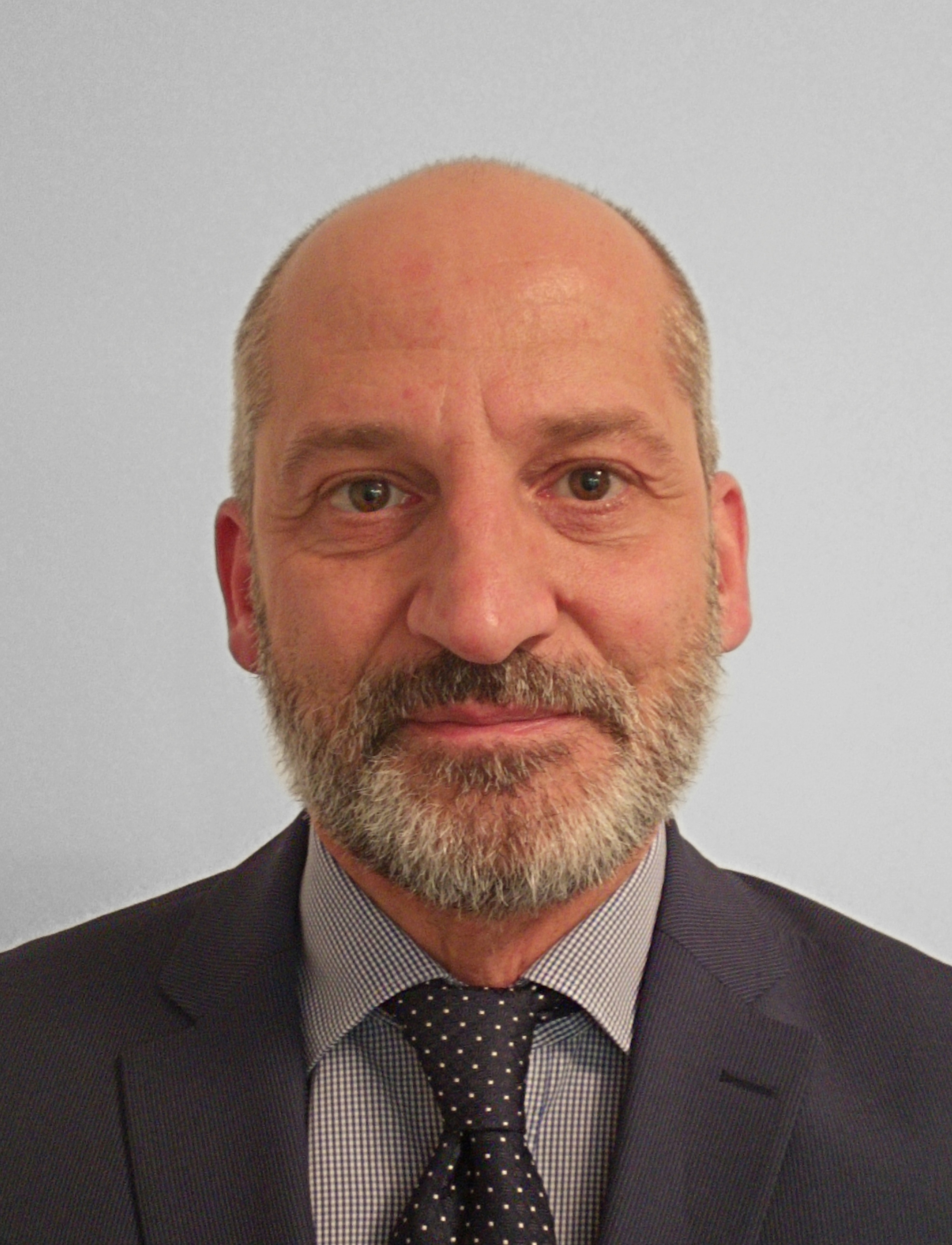}}]{PANAGIOTIS TRAKADAS}~(MEng, Ph.D) received his Diploma, Master Engineering, Degree in Electrical $\&$ Computer Engineering and his Ph.D. from the National Technical University of Athens. His main interests include 5G/6G technologies, such as O-RAN, NFV and NOMA, Wireless Sensor Networks, and Machine Learning based Optimization Techniques for Wireless Systems and Maritime Communications. 

He is currently an Associate Professor at the Department of Ports Management and Shipping, in National and Kapodistrian University of Athens and has been actively involved in many national and EU-funded research projects. He has published more than 170 papers in magazines, journals, books and conferences and a reviewer in several journals and TPC in conferences.
\end{IEEEbiography}

\vfill

\end{document}